# Causal-order superposition as an enabler of free will

Salvador Malo[*]

*October 4, 2016*

**Abstract.** It is often argued that bottom-up causation under a physicalist, reductionist worldview precludes free will in the libertarian sense. On the one hand, the paradigm of classical mechanics makes determinism inescapable, while on the other, the leading models that allow a role for quantum effects are noncommittal regarding how conscious agents are supposed to translate indeterminacy into self-formed choice. Recent developments, however, not only imply that self-formed decisions are possible, but actually suggest how they might come about. The cornerstone appears to be causality superposition rather than quantum-state entanglement, as is usually assumed, and the natural arena for applying these developments is (perhaps ironically) a framework that was built without any consideration for quantum effects.

The libertarian notion of free will, specifically the interpretation denoted as (freedom$_2$) in [3][†], tends to come under fire for assuming the possibility of actions that are self-forming, *i.e.* neither random nor determined by pre-existing conditions. Under the paradigm of classical physics, determinism follows from bottom-up causation, and self-forming actions are a nonstarter. Quantum mechanics, on the other hand, because it allows actions to be triggered by random microscopic fluctuations, has often been interpreted (*e.g.* in [6]) as merely replacing the yoke of determinism by that of chance, which is hardly an improvement. This line of criticism has gained credence because the strongest proposals that call attention to quantum indeterminacy as an enabler of free will (notably [5,10]) do not explain to the satisfaction of naysayers how the harnessing of indeterminacy takes place, in other words what makes decoherence a choice as opposed to a mere event.

**Causal Order and Reductionism**

When the bottom flaps of a cardboard box are tucked under their neighbors in an overlapping pattern, the resulting floor can be rigid enough to carry lightweight objects without the help of tape. By denoting the set of flaps of a square box as $F = \{F_1, F_2, F_3, F_4\}$, it is clear that its strength is a property of $F$ but not of any single $F_i$ in isolation. (If the flaps could voice an opinion, they might claim their individual role to be merely that of passing weight on to a neighbor, thus making no effort of their own.)

Ignoring the subtleties of mechanics and following this abstraction to its conclusion, it could be said that $F$ as a whole (rather than any $F_i$ alone) is what prevents the object from falling. Moreover, in the causal chain that begins with an object being placed inside the box and ends with the object either being supported or not, $F$ is clearly a middle link. But any attempt to increase the granularity of causation by zooming into the elements of $F$ runs into trouble, for it is impossible to select one particular flap $F_i$ as the first link within that subset of the causal chain. In short, causal order within $F$ itself is indefinite.

---

[*] salvador.malo.g@gmail.com
[†] To distinguish it from the harder-to-defend (freedom$_3$)

From a physicalist, bottom-up causation viewpoint the above abstraction is of course untenable, because the floor of the box is a macroscopic object of which $F$ is but an imperfect description. The box comprises a vast number of material particles that are acted upon by forces that only in concert grant rigidity to $F$ and prevent the object from falling through. A careful (even if impractical) dissection of all the microscopic elements involved should in principle reveal a first link in the causation chain, whether a first collision of a subatomic particle of the dropped object against another particle from the cardboard surface, or a random fluctuation that unleashes a sequence of events that ends in the object falling or staying put.

"So what?" might claim the engineer, for whom microscopic fluctuations are irrelevant as long as the statistical average of the forces involved allows the object to be supported for a reasonable amount of time. (Nobody cares whether a box lies at the mercy of particle collisions as long as it fulfills its intended purpose.) But when human brains and other ostensibly conscious agents are subject to the same scrutiny, such dismissive attitudes become unpalatable, and the resulting discomfiture has led even to the suggestion that the concept of causation must be re-examined [4].

**Conscious Causality**

Suppose that a 3-member committee $M = \{M_1, M_2, M_3\}$ generates approval/rejection rulings $R$ on proposals $P$ based either on immediate voting or on a deliberation process $D$, assumed to take place behind closed doors and without communication with the outside world.

Immediate voting is algorithmic and generates $R$ as a simple average of member votes, which in turn reflect the pre-existing views of committee members. In contrast, $D$ allows information exchange between the $M_i$ to influence each other's views. $R$ remains undetermined until $D$ ends, even if individual views are polled before $D$ begins. Moreover, the order in which $M_i$ can influence $M_j$ is indefinite (as anyone knows who has participated in deliberations). Of course, this is only because individual members play the role of homunculi within $M$. But it will be useful to summarize the setup in terms that can be applied later:

1. Since each $M_i$ is assumed to be autonomous, the order in which they exchange information is indefinite, so $R$ is undetermined until it is produced, even if the initial states of all $M_i$ are known
2. But $R$ isn't random, for it depends on the views of $M$'s elements.

$M$ is therefore irreducibly the free causal agent of outcome $R$. However, the translation of this metaphor to the case of an individual mind is problematic, because the autonomy of a mind's elements is questionable, and because it isn't immediately clear what the equivalent of $D$ should be within the context of a physical brain. In order to reformulate it for such a context, several obstacles must be overcome.

**Delimiting the Agency**

First, there is the question of delimitation, or the extension of the physical agent. Is it the person, the person's brain (excluding the skull but including the spinal cord and all its nerve endings), the neocortex, a specific network within a part of the brain, a set of microtubules—or none of the above? Although its explanatory power as a model of consciousness is strongly contested, Integrated Information Theory (IIT) goes a long way towards characterizing decision agents, so for the purpose of this discussion an agent $A$ shall be assumed to be an *integrated system* as defined in [7]. IIT's definition is a formal abstraction that circumvents the issue of $A$'s physical constitution, but it identifies which states of a system can play the roles of cause and effect, and excludes everything else from the discussion (a much-welcome simplification). A scaled-down version of an example from [7] is drawn below, with input $I$, output $O$, elements $A_i$ and connections that are either excitatory (solid) or inhibitory (dotted). Internal feedback loops are one of its essential features. In order to forestall the threat of homunculi, the elements $A_i$ are assumed to be nothing more than logic gates. Can $A$ be characterized so that its properties are similar to those of $M$ in the committee example?

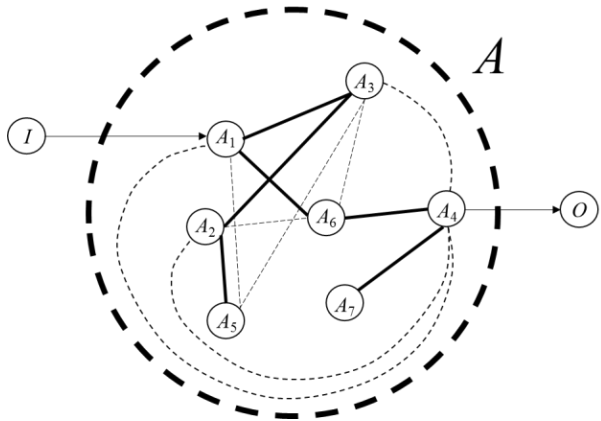

If $A$ is assumed to behave classically, then for elements $A_i$ in known initial states, and for a given input $I$, the production of $O$ follows a deterministic causal chain that begins with $I$ and terminates with the first output $O$ (the persistence of feedback within $A$ may allow $O$ to be followed by more outputs, but that would merely extend the deterministic chain).

**Shielding $A$ from Chance**

If $A$ is assumed to behave quantum-mechanically, then several additional prospects arise, not all of them necessarily good. There is, to begin with, a possible susceptibility to quantum fluctuations, which in order to be useful rather than a nuisance must be exploited by the agent rather than by the environment. In the absence of such advantage, it is clearly preferable for the $A_i$ to be impervious to off-network noise, thus preventing unwelcome microscopic randomness from being amplified to macroscopic scale. Within-network quantum effects, however, are another matter, and it is ironical that IIT is formulated entirely in classical terms, for it is ideally suited for quantum interpretations of causality.

The next prospect is superposition. But what sort of superposition, if any, increases $A$'s freedom of action? The literature's favored candidate (quantum state entanglement) may in fact be a red herring. In the case of the fictitious committee $M$ described earlier, it isn't helpful for the opinions of individuals to be entangled (so that *e.g.* measuring $M_2$'s view becomes redundant once $M_1$'s is known). That would turn $M$ into a condensate—a partisan, rubberstamping, ineffectual blob. The chief virtue of deliberation is that it prevents regrets arising from uninformed voting. Stated otherwise, $D$ is meant to improve decision-making, not to render individual views indistinguishable.

Returning to $A$, its appeal as an agency is the richness of its internal network, the complexity of which is worth exploiting rather than smearing out. In the example above, $A_1$ is by design the first element to react to $I$. Also by design, once $A_4$ is first acted upon, measurable output is generated which constitutes a causal endpoint for external observers, even if internal feedback subsequently allows additional signals to emerge as $O$. IIT shows in [7] that such networks are functionally equivalent to deterministic "zombie" systems. But this is only because IIT's conceptual approach is formulated without any reference to quantum effects. If the elements $A_i$ were allowed to communicate in no definite order, yet still according to the rules of the network (even if performing *known* operations, or the equivalent of an $M_i$ speaking its mind), then $A$ would remain the irreducible cause of $O$ without its output being either smeared out to trivial randomness or cranked out Turing-like. What is called for isn't state entanglement, but causal superposition, where the $A_i$ are defined but the order in which they communicate is not. As it happens, causal superposition is not only possible in theory [8] but has been implemented in practice [9]. But, why is it useful?

**Shielding $A$ from Determinism**

As the diffraction pattern from a double-slit experiment famously shows, a photon behaves differently when presented with several alternative paths than when presented with only one. The alternative in question is the path through the other slit, and the different behavior is the impact spot which (in the aggregate) builds up a diffraction pattern. One interpretation of this effect is that a photon's actual impact is influenced by the alternative trajectories it might have taken. What *might have been* influences what *in fact was*. For a conscious being habituated to constant what-if considerations, this is nothing special, but for a single photon's behavior it is perplexing, to say the least.

When speaking of causal-order superposition instead of state superposition, the alternative in question involves cause-effect reversal rather than a state entanglement. For a process in causal-order superposition, two distinct events are in the past of each other. Not surprisingly, the mathematics are nontrivial, but a precise description is given in [8], which shows that this form of superposition cannot be described by a probabilistic mixture of processes with definite causal order. Moreover, the framework avoids paradoxes such as those stemming from closed time-like curves.

The implication is that when the signaling over a network shared by $A_1$ and $A_2$ is in causal-order superposition and a qubit $I$ is sent from one element to another, then even when information flows in just one sense (say, from $A_1$ to $A_2$), the mere possibility that it "might have" gone in the opposite sense is enough for $A_1$ to make a better guess of $A_2$'s measurement (of $I$) than in the absence of causal superposition. Stated in terms of informatics, the outcome of a logic-gate's operation on an undetermined input can be influenced by the might-have-been operation of another logic gate in its causal future, had it been in its causal past. This allows better-informed outputs $O$ to be generated by a system *A* than would be produced if its elements operated under a definite causal order. The situation is manifestly different from that of classical feedback, where ordinary causality merely allows an operation performed by $A_1$ to be overwritten at a later time (based on $A_2$'s classical feedback), but doesn't allow $A_2$ to influence the *first* signal of $A_1$!

The framework has been extended in [1] to show that absence of causal order allows perfect signaling between 3 parties the first time around (*i.e.* perfect agreement of their first signals). Even more complex examples will surely be developed as the logic and the consequences of the framework are further explored. But the work done in [9] already proves that the above isn't mere speculation: physical systems can be experimentally set up so that their individual elements do not (even in principle) exchange information in a particular time order. If *A* is set up in such a way (and if science has built a simple example, why wouldn't nature have built more elaborate ones?) this means that *A*'s decisions cannot not be determined (even in principle) by full knowledge of *I* and of the states of the elements $A_i$ at any time before *O* is produced. After input *I* is fed into a system *A* whose elements communicate in a causally indefinite order, its output *O* can be claimed to be caused by *A*, and might arguably be said to be "influenced" by *I* (which would require yet another definition), but is emphatically not *determined* by the pre-existing state of elements $A_i$ and/or input *I*.

Despite its classical formulation, IIT contributes more than just an abstract description of purportedly conscious agents. One of its key conclusions is that "experiences" (defined in a precise way in [7], but meaning *A*'s internal representations of *O*) are intrinsic properties of integrated systems in a given state. This conclusion relies merely on the ability of IIT to describe the neural mechanisms of subjective perception, not on its validity as a model of consciousness in a more general sense (and even less on whether or not it addresses the Hard Problem [2] or is subject to fading qualia arguments). What IIT explicitly shows is how experiences are self-generated by *A* alone and in a manner that is self-referential to *A* alone. To wit, *A* doesn't have to generate any measurable output, nor does it need to consult any additional[‡] input in order to form an internal representation of *O*.

This deserves emphasizing, for it spells out, in the context of *A,* an equivalent of the committee's deliberation *D*. The feedback or re-entry required of integrated systems in [7] paves the way for *A* to experience a given *O* without reference to its resultant output. This is crucial, for if *O* had to be written out in order for *A* to experience it, then it would already be too late for the choice to be examined before being made. The examples outlined in [1,8] and explicitly built in [9] are all defined in terms of co-operative goal achievement, the likelihood of which is maximized under indefinite causal order. The goal in those examples is arbitrarily imposed by the experimenters and (by necessity) rather simple, but in more complex cases could be defined in terms of *A*'s own subjective standards, based for example on the stored representations of previous experiences. Causal-order superposition would then lead to the instant production of an output *O* satisfying such internally-defined goal, which would also be internally representable (perhaps later, as a classically stored memory).

To summarize, causal-order superposition applied to an integrated system *A* defined as in [7] but subject to the requirements of [9] would allow *A* to carry out actions that are formed entirely within

---

[‡] Additional because input *I* was earlier assumed have already been fed into *A*.

*A,* based on *A*'s experience of possible alternatives, and neither random nor deterministic. Proof may yet come in experimental form (as it always should), but what to call this if not libertarian free will?